\documentclass[conference]{IEEEtran}
\IEEEoverridecommandlockouts

\usepackage{cite}
\PassOptionsToPackage{numbers}{natbib}
\usepackage{amsmath,amssymb,amsfonts}
\usepackage{amssymb}
\usepackage{algorithm}
\usepackage{algorithmic}
\usepackage{booktabs}
\usepackage{graphicx}
\usepackage{textcomp}
\usepackage{xcolor}
\usepackage{hyperref}
\usepackage{soul}

\def\BibTeX{{\rm B\kern-.05em{\sc i\kern-.025em b}\kern-.08em
    T\kern-.1667em\lower.7ex\hbox{E}\kern-.125emX}}
\begin{document}

\title{Quantum Multimodal \\ Contrastive Learning Framework \\
\thanks{This paper has been accepted by the 2025 International Conference on Acoustics, Speech, and Signal Processing (ICASSP 2025).}
}

\author{
    \IEEEauthorblockN{Chi-Sheng Chen\IEEEauthorrefmark{1}, 
     Aidan Hung-Wen Tsai\IEEEauthorrefmark{1}, 
     Sheng-Chieh Huang\IEEEauthorrefmark{2}}
    
    \IEEEauthorblockA{\IEEEauthorrefmark{1}Neuro Industry, Inc., CA, USA \\
    m50816m50816@gmail.com, \{michael, aidan\}@neuro-industry.com}

    \IEEEauthorblockA{\IEEEauthorrefmark{2}Department of Electronics and Electrical Engineering, National Yang Ming Chiao Tung University, Hsinchu, Taiwan\\
    shengchiehhuang@nycu.edu.tw}
    
}

\maketitle
\begin{abstract}
In this paper, we propose a novel framework for multimodal contrastive learning utilizing a quantum encoder to integrate EEG (electroencephalogram) and image data. This groundbreaking attempt explores the integration of quantum encoders within the traditional multimodal learning framework. By leveraging the unique properties of quantum computing, our method enhances the representation learning capabilities, providing a robust framework for analyzing time series and visual information concurrently. We demonstrate that the quantum encoder effectively captures intricate patterns within EEG signals and image features, facilitating improved contrastive learning across modalities. This work opens new avenues for integrating quantum computing with multimodal data analysis, particularly in applications requiring simultaneous interpretation of temporal and visual data.
\end{abstract}



\section{Introduction}
Knowing how the human brain processes visual information is a challenging task that connects neuroscience and artificial intelligence (AI). The brain uses a complex system of neural activities across different regions to analyze what we see. This understanding has helped in creating deep learning methods, like convolutional neural networks (CNNs). Further research into how the brain processes visual information in real-life situations could drive new AI advancements.

Recent studies have improved our knowledge by observing brain activity through various techniques. Electroencephalography (EEG) is particularly useful because it is non-invasive, portable, and can capture fast changes in brain activity. EEG gives us a unique way to study how the brain processes and recognizes visual information as it happens in real time. Recently, EEG has seen many interesting applications of machine learning, such as predicting treatment outcomes in psychiatry \cite{li_chen_mdd_ai_2023}, identity recognition \cite{mindid_2018}, and robotic control \cite{eeg_robo_2016}, etc. These applications rely on the clear and real-time temporal information provided by EEG.



In recent years, contrastive learning has become a powerful tool in self-supervised learning, especially for multimodal data. By associating related pairs from different modalities, it has achieved impressive results in tasks like image-text alignment and audio-visual correspondence. However, classical approaches struggle with high-dimensional, complex data such as EEG, making image decoding from EEG signals challenging due to the low quality and unpredictability of the signals. Overcoming these challenges is essential for advancing visual processing and improving EEG-based image decoding and brain-computer interfaces (BCIs). Earlier attempts to decode images from EEG data were limited by small datasets, restricting model generalization. While recent efforts with larger datasets and rapid image presentations have made some progress, they are still constrained by classical neural networks. These limitations underscore the need for EEG processing techniques that align better with the brain’s mechanisms and the unique properties of EEG data.

Although quantum methods have yet to surpass traditional approaches, recent findings in \cite{chen_qeegnet_2024} suggest that quantum encoders hold promise for EEG processing. Inspired by these results, we developed a Quantum Multimodal Contrastive Learning Framework (QMCL), an exploratory approach that leverages quantum computing’s unique properties, such as superposition and entanglement, to enhance learning across different data modalities. By integrating Variational Quantum Circuits (VQCs) with classical neural networks, our approach explores a broader solution space, potentially enabling more efficient and effective multimodal representation learning. While still in its early stages, this hybrid quantum-classical framework shows promise for surpassing the limitations of classical methods and unlocking new possibilities for complex applications like EEG processing.

Our QMCL framework offers a novel approach to tackling the complexities of EEG data, contributing the following advancements:

\begin{itemize}
\item To the best of our knowledge, this is the first work to utilize a VQC architecture for self-supervised cross-modal contrastive learning.
\item We propose a Quantum Multimodal Contrastive Learning Framework (QMCL) and apply it to a zero-shot classification task involving EEG and images.
\item We are the first to use a quantum algorithm on one of the largest EEG-image datasets, setting a new benchmark for future quantum contrastive learning research.
\item We specifically propose four hybrid quantum-classical time-series encoders tailored for QMCL and introduce the Quantum-CLIP-ViT image encoder for this task.
\end{itemize}

\begin{figure}
    \centering
    \includegraphics[width=1\linewidth]{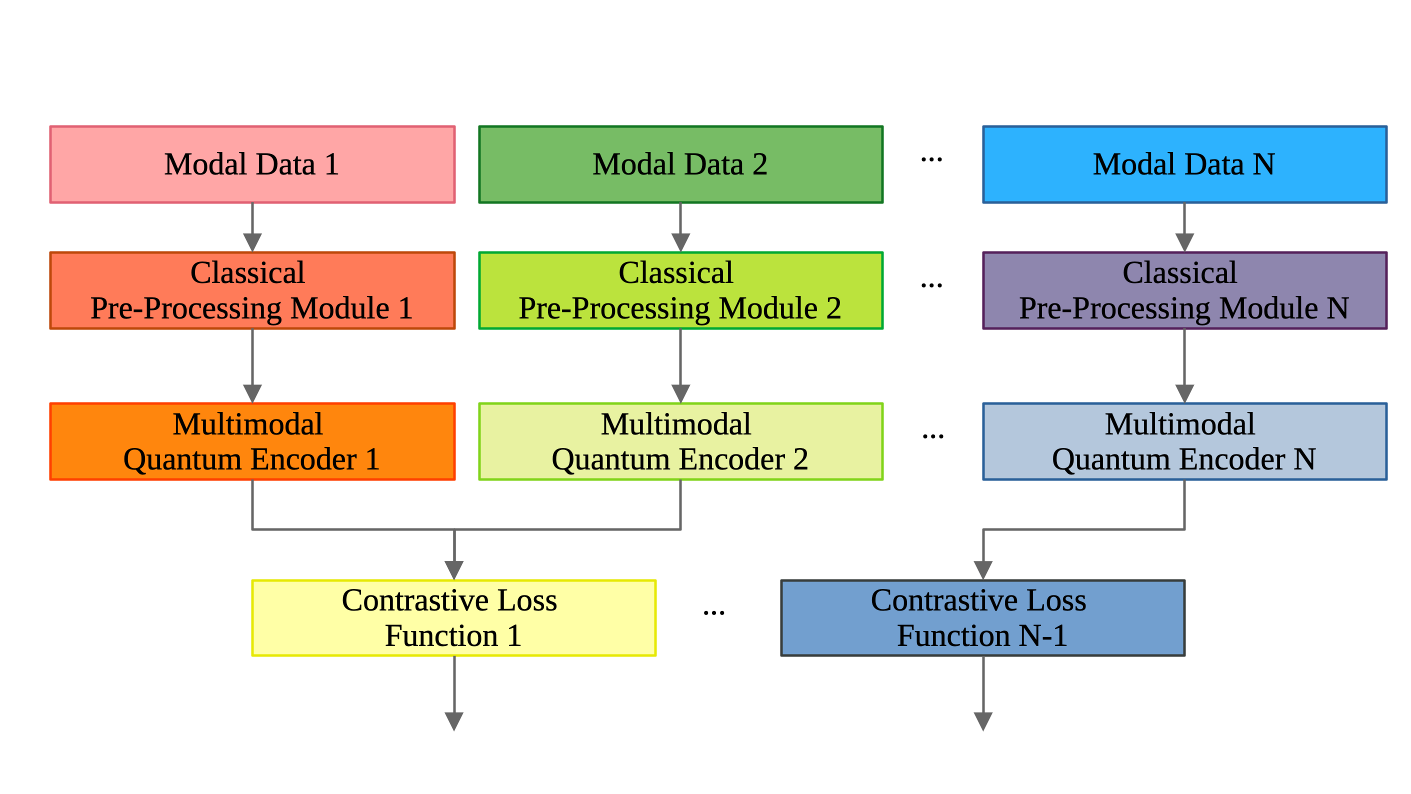}
    \caption{Overview of the Multimodal Quantum Contrastive Learning (QMCL) framework. Multiple modal data sources are first processed through classical pre-processing modules. The processed data are then encoded using multimodal quantum encoders, with each encoder corresponding to a specific data modality. The encoded representations are compared using contrastive loss functions to align the features across different modalities.
    }
    \label{fig:enter-label}
\vskip -0.2in
\end{figure}
\begin{figure}
    \centering
    \includegraphics[width=1\linewidth]{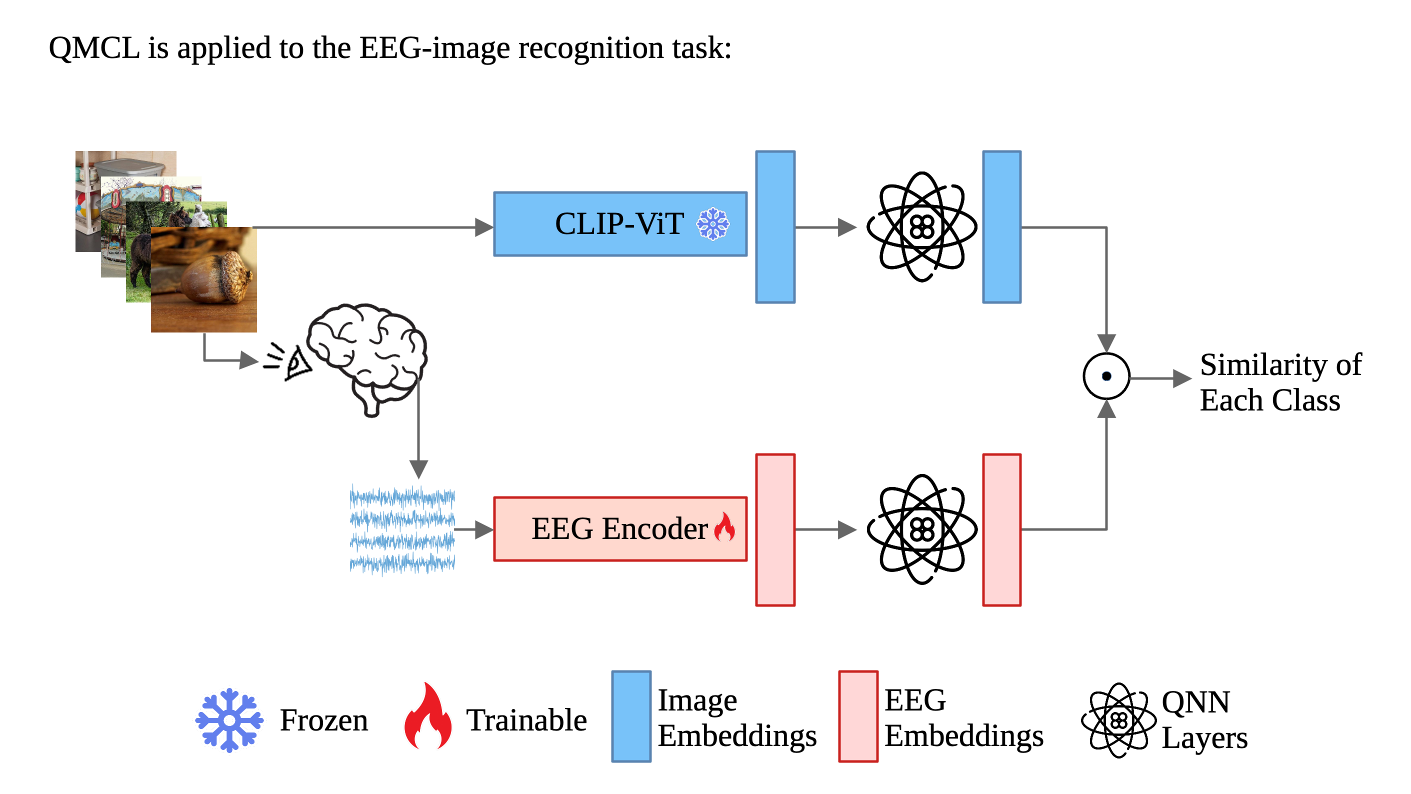}
    \caption{Application of QMCL to EEG-image recognition. Image embeddings are generated using a frozen CLIP-ViT model, while EEG embeddings are produced by a trainable EEG encoder. Both embeddings are processed through quantum neural network (QNN) layers to compute the similarity of each class, enabling the zero-shot recognition task.}
    \label{fig:qmcl_eeg_img}
\vskip -0.3in
\end{figure}

\section{Related Work}
\vspace{-0.1in}
\subsection{Classical Contrastive Learning}
\vspace{-0.1in}

In multimodal contrastive learning, CrossCLR \cite{zolfaghari_zhu_gehler_brox_2021} extends the approach to cross-modal representation learning by leveraging correlations between different modalities. CLIP \cite{clip_2021} marks a major breakthrough, learning text-image correspondences in an unsupervised setting. Similarly, ALIGN \cite{align_2021} aligns images and text within a shared embedding space through large-scale pre-training, showcasing the practical potential of multimodal contrastive learning. These models demonstrate the evolution of contrastive learning from unimodal to multimodal applications, underscoring its versatility across data modalities. From SimCLR \cite{simclr_2020} and MoCo \cite{moco_2019} to CLIP and ALIGN, contrastive learning has emerged as a key technique in unsupervised learning, driving advancements across various fields. \cite{chen2024MUSE} has tried to do EEG-image contrastive learning with transformer models, and the work is the state-of-the-art of the task.   

\vspace{-0.1in}
\subsection{Quantum Machine Learning on Time-Series Data}
\vspace{-0.1in}
Recent efforts have extended quantum machine learning (QML) to time-series data, aiming to leverage quantum information properties to either reduce model parameters while maintaining performance or improve data representations. Hybrid quantum-classical VQC algorithms for time-series preserve the neural network’s ability to extract temporal and spatial features while mapping learned representations to a Hilbert space via quantum layers. \cite{ts_forecast_qml_2022} presents two hybrid architectures inspired by multilayer perceptrons (MLP), evaluated on univariate (Mackey-Glass, USD-to-euro) and multivariate (Lorenz attractor, Box-Jenkins Gas Furnace) time series. These models performed competitively with MLP, CNN, and LSTM models of similar complexity, suggesting that integrating quantum elements can enhance classical models. 
\cite{samuel_qlstm_2022} introduced a hybrid quantum-classical LSTM model (QLSTM) that effectively learns temporal data, sometimes converging faster or achieving better accuracy than classical LSTMs. The variational approach in QLSTM reduces qubit and circuit depth requirements, making it suitable for noisy intermediate-scale quantum (NISQ) devices. For EEG applications, \cite{chen_qeegnet_2024} proposed Quantum-EEGNet (QEEGNet), a hybrid model integrating quantum computing with the classical EEGNet \cite{lawhern_solon_waytowich_gordon_hung_lance_2018} architecture. QEEGNet enhances EEG encoding by capturing intricate patterns in EEG data, demonstrating the significant potential of quantum-enhanced neural networks in EEG analysis and paving the way for further research and practical applications.
\vspace{-0.1in}
\subsection{Quantum Contrastive Learning}
\vspace{-0.1in}
Recent advances in quantum machine learning have primarily focused on unimodal data, enhancing tasks like image classification and time series prediction. All the following works concentrate on single-modality data without addressing the complexities of multimodal learning. \cite{qml_cons_2022} introduces a hybrid quantum-classical architecture for self-supervised visual representation learning, demonstrating the numerical advantages of small-scale QNNs and their comparable performance to classical models on real quantum devices. \cite{fu_sqcl} proposes a model that combines Supervised Contrastive Learning (SCL) with VQC and uses Principal Component Analysis (PCA) for dimensionality reduction, achieving high accuracy in medical image analysis, especially with a small number of qubits. \cite{Q-SupCon} presents a quantum-enhanced supervised contrastive learning model addressing data scarcity, showing significant accuracy in image classification tasks with limited labeled data and stable performance on real quantum devices. \cite{qclr_lstm_2024} proposes a quantum-enhanced self-supervised learning framework for mental health monitoring, utilizing a quantum-enhanced LSTM encoder that significantly outperforms traditional models in predictive capability, achieving high F1 scores across multiple datasets.
Despite these advancements, all existing works are confined to unimodal data. In contrast, our work introduces Quantum Multimodal Contrastive Learning (QMCL), the first approach to leverage quantum computing for learning across multiple data modalities. This breakthrough allows for richer and more complex representations, opening new avenues for applications that require the integration of diverse data sources.

\begin{table*}[t]
    \caption{Comparison of the top-1 accuracy results of QSTConv, QSTConv-GA, QNervFormer and QNervFormer-GA on the ThingsEEG test Dataset for 200-way zero-shot classification.}
    \centering
    \begin{tabular}{cccccccccccc}
                &  Subject 1 & Subject 2 & Subject 3 & Subject 4 & Subject 5 & Subject 6 & Subject 7 & Subject 8 & Subject 9 & Subject 10 & Average\\
       QSTConv   &  3.2\% & 2.1\% & 4.5\% & 3.5\% & 2.0\% & 1.8\% & 3.4\% & 4.6\% & 3.8\% & 6.0\% & 3.5\% \\
       QSTConv-GA  & 4.2\% & 3.0\% & 3.3\% & 6.0\% & 2.6\% & 3.6\% & 4.9\% & 4.4\% & 3.9\% & 3.9\% & 4.0\% \\
       QNervFormer  & \textbf{4.3\%} & 4.1\% & \textbf{5.6\%} & 5.6\% & 4.3\% & 4.4\% & 5.5\% & 4.8\% & \textbf{5.9\%} & \textbf{6.6\%} & 5.1\% \\
       QNervFormer-GA & 3.9\% & \textbf{4.2\%} & 5.4\% & \textbf{6.1\%} & \textbf{4.4\%} & \textbf{4.5\%} & \textbf{6.4\%} & \textbf{5.8\%} & 5.6\% & 6.1\% & \textbf{5.2\%}\\
    \end{tabular}
    \label{tab:qmcl_top1}
\vskip -0.2in
\end{table*}

\begin{table*}[t]
    \caption{Comparison of the top-5 accuracy results of QSTConv, QSTConv-GA, QNervFormer and QNervFormer-GA on the ThingsEEG test Dataset for 200-way zero-shot classification.}
    \centering
    \begin{tabular}{cccccccccccc}
                & Subject 1 & Subject 2 & Subject 3 & Subject 4 & Subject 5 & Subject 6 & Subject 7 & Subject 8 & Subject 9 & Subject 10 & Average\\
       QSTConv  & 12.6\% & 9.6\% & 16.8\% & 12.4\% & 7.6\% & 8.9\% & 13.3\% & 20.0\% & 16.3\% & 19.7\% & 13.7\% \\
       QSTConv-GA  & 16.7\% & 13.4\% & 15.0\% & \textbf{24.4\%} & 11.4\% & 16.7\% & 16.6\% & 20.0\% & 13.5\% & 20.3\% & 16.8\% \\
       QNervFormer  & \textbf{18.0\%} & \textbf{18.6\%} & \textbf{21.9\%} & 23.1\% & \textbf{15.5\%} & \textbf{20.2\%} & 20.6\% & 19.8\% & \textbf{20.3\%} & 23.7\% & 20.2\%\\
       QNervFormer-GA  & 17.4\% & 17.7\% & 20.2\% & 22.4\% & 14.5\% & 19.5\% & \textbf{23.0\%} & \textbf{24.5\%} & 19.3\% & \textbf{24.8\%} & \textbf{20.3\%}\\
    \end{tabular}
    \label{tab:qmcl_top5}
\vskip -0.2in
\end{table*}
\section{Methodology}


The proposed Quantum Multimodal Contrastive Learning (QMCL) framework is designed to leverage the advantages of quantum computing for learning rich and complex representations across multiple data modalities. Unlike previous quantum-enhanced contrastive learning methods that focus exclusively on a single modality, QMCL introduces a hybrid quantum-classical architecture capable of handling and integrating data from diverse sources such as images, text, and time series. 
As the application example of QMCL in this work, we propose a QMCL framework that integrates various quantum-enhanced EEG encoders with a quantum CLIP-ViT for multimodal contrastive learning. The core idea is to leverage the power of quantum computing to encode both EEG signals and images into high-dimensional quantum states and perform contrastive learning between these modalities. The architecture is designed to maximize the similarities between corresponding EEG-image pairs while pushing apart non-corresponding pairs, thus enabling effective multimodal representation learning. The whole architecture illustrated in Fig~\ref{fig:qmcl_eeg_img}.
\vspace{-0.1in}
\subsection{Image Processing with QCLIP-ViT}
\vspace{-0.1in}
For image data, we employ the Quantum CLIP-ViT (QCLIP-ViT). The image is first divided into patches, which are flattened and linearly projected. Positional embeddings are added to preserve spatial information, followed by a series of transformer layers, including multi-head self-attention and feed-forward layers. The processed image embeddings are then passed through quantum encoding layers, which map the classical embeddings into quantum states, preparing them for contrastive learning with the EEG embeddings.

\vspace{-0.1in}
\subsection{EEG Processing with Quantum Encoders}
\vspace{-0.1in}
We implement several variations of quantum encoders tailored for EEG data which are inspired from our previous work \cite{chen2024MUSE}, each designed to capture different aspects of the time-series signals:
\subsubsection{QNervFormer}
\begin{figure}
    \centering
    \includegraphics[width=1\linewidth]{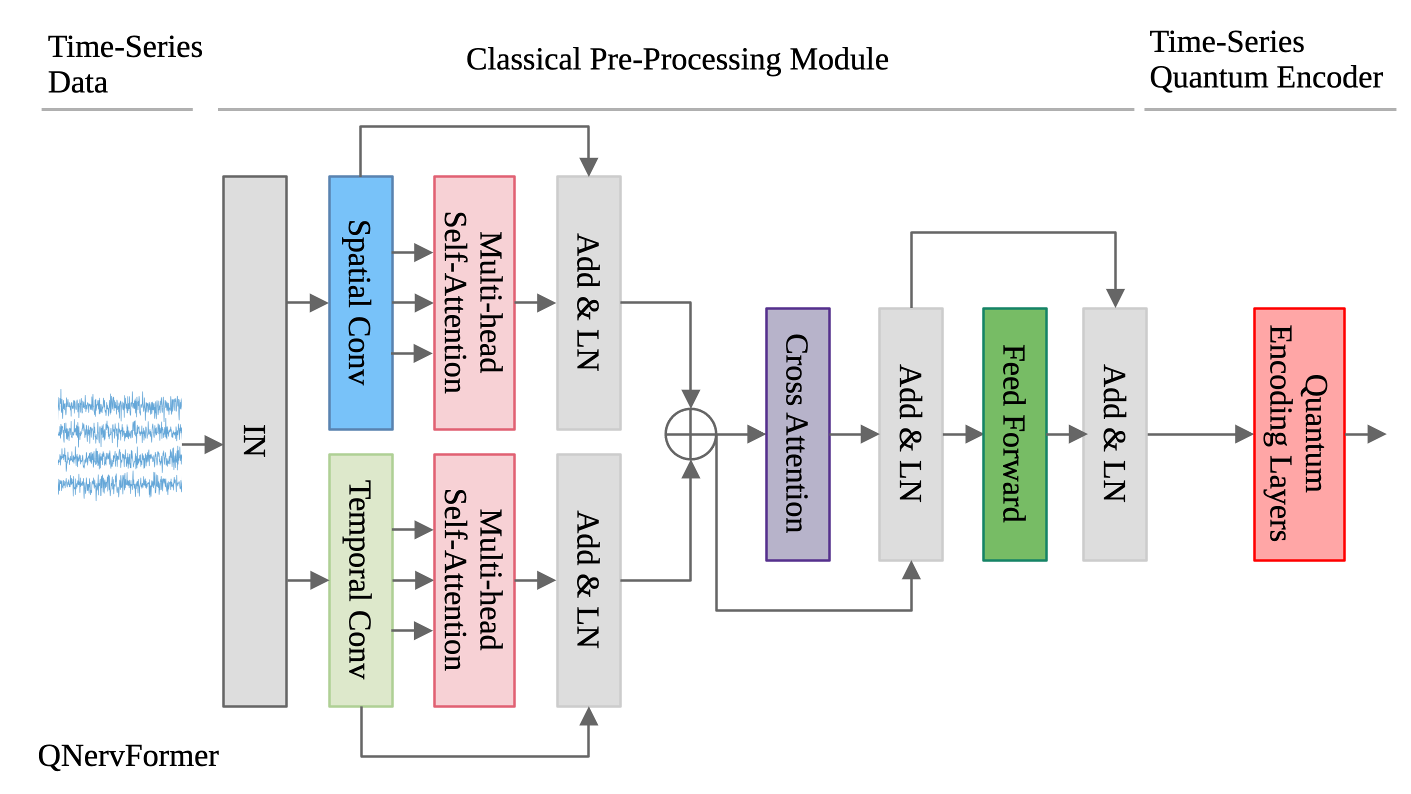}
    \caption{Architecture of the QNervFormer model for time-series data encoding. The model processes time-series data through spatial and temporal convolutions, followed by multi-headed self-attention mechanisms. The final quantum encoding layers enhance the representation of the input data for downstream tasks. Where IN and LN note instance and layer normalization, respectively.}
    \label{fig:qnf}
\vskip -0.3in
\end{figure}

Combines spatial and temporal convolutional layers followed by multi-head self-attention and cross-attention layers, ultimately passing the features through quantum encoding layers. The model shows in Fig~\ref{fig:qnf}.
\subsubsection{QNervFormer-GA} 
\begin{figure}
    \centering
    \includegraphics[width=1\linewidth]{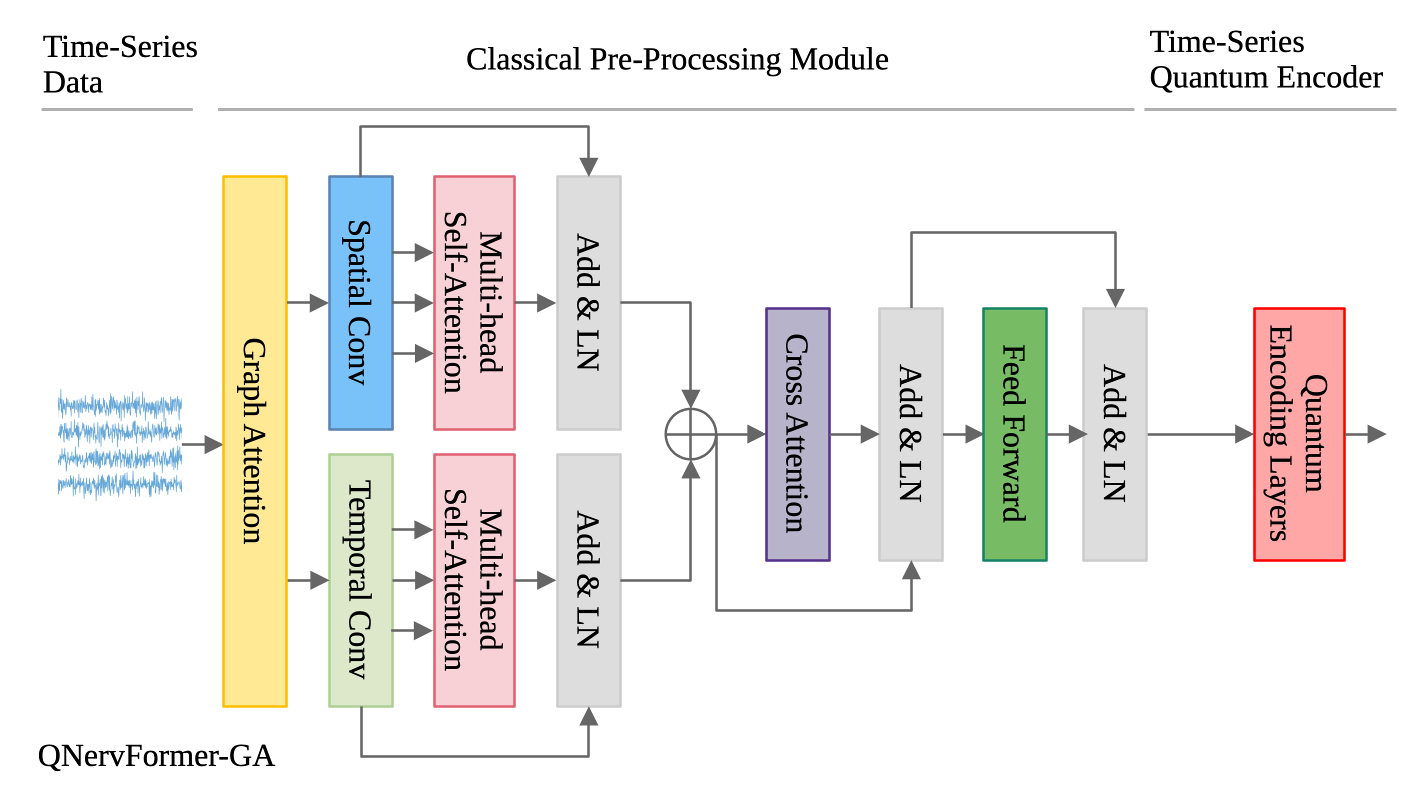}
    \caption{Architecture of QNervFormer-GA, an extension of QNervFormer with integrated graph attention.}
    \label{fig:qnfga}
\vskip -0.3in
\end{figure}

Extends QNervFormer by incorporating graph attention mechanisms \cite{velivckovic2017graph}, enhancing the spatial relationships between EEG channels. The model shows in Fig~\ref{fig:qnfga}.
\subsubsection{QSTConv}
Inspired by EEGNet, the model first do spatial than do temporal convolutions, followed by quantum encoding layers.
\subsubsection{QSTConv-GA}
\begin{figure}
    \centering
    \includegraphics[width=1\linewidth]{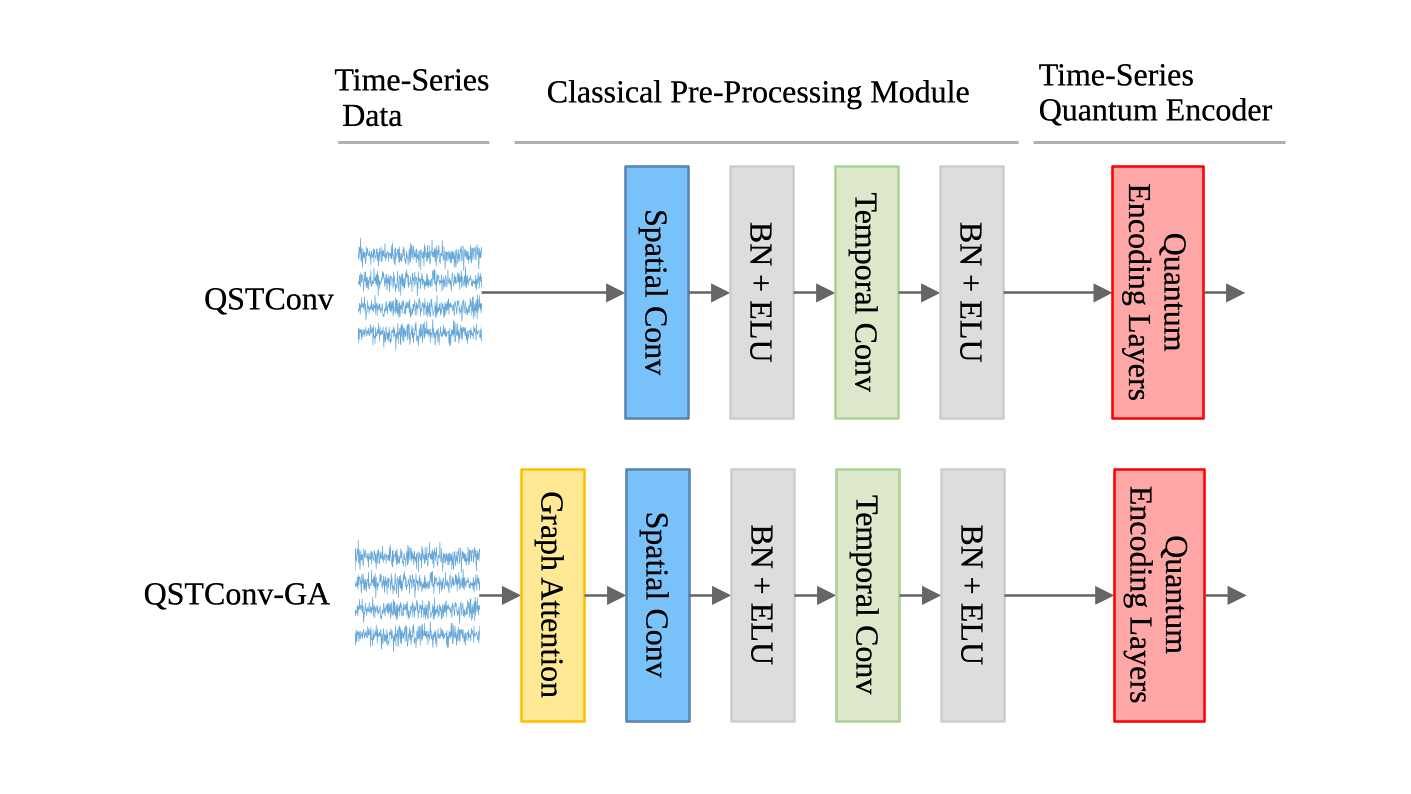}
    \caption{Comparison of QSTConv and QSTConv-GA architectures for time-series data encoding. The QSTConv model applies spatial and temporal convolutions with batch normalization (BN) and ELU activation, followed by quantum encoding layers. The QSTConv-GA model introduces graph attention as an additional feature extraction mechanism, integrating it into the pre-processing pipeline before quantum encoding.}
    \label{fig:qstcov}
\vskip -0.3in
\end{figure}

Utilize spatial and temporal convolutions, with the latter also incorporating graph attention layers, followed by quantum encoding layers. The QSTConv series models are illustrated in Fig~\ref{fig:qstcov}.
Each EEG encoder processes the time-series data through classical pre-processing modules, which include convolutional and attention layers, before encoding the output into quantum states. These quantum-encoded EEG features are then aligned with the image embeddings for contrastive learning. 


This superposition allows the quantum encoding layer to encode and process more information than a classical layer of comparable size.
\vspace{-0.1in}
\subsection{Quantum Encoding Layer}
\vspace{-0.1in}
The quantum encoding layer integrates quantum computing capabilities into classical neural networks. By embedding a quantum circuit within a classical deep learning model, the quantum encoding layer leverages the unique properties of quantum mechanics—such as superposition and entanglement—to enhance the performance and capabilities of machine learning models. This layer utilizes a parameterized quantum circuit, where classical EEG features are mapped to a set of qubits $\mathbf{x} \in \mathbb{R}^{n_{\text{qubits}}}$, with trainable weights $\mathbf{w} \in \mathbb{R}^{n_{\text{layers}} \times n_{\text{qubits}}}$.

\textbf{1. Input Encoding:} Each qubit $q_i$ is rotated based on the input data $x_i$ using the rotation-Y (\text{RY}) gate:
\begin{equation}
    RY(x_i) = \exp\left(-i \frac{x_i}{2} \sigma_y\right),
\end{equation}
where $\sigma_y$ is the Pauli-Y matrix.

\textbf{2. Parameterized Layers:} The quantum encoding layer uses a ring pattern of controlled-NOT (CNOT) gates. 
The ring pattern involves applying CNOT gates to entangle each qubit with its neighbor, forming a closed loop. For each layer $l$ in $n_{\text{layers}}$, the CNOT gates are applied as follows:
\begin{equation}
    \text{CNOT}(q_i, q_{(i+1) \mod n_{\text{qubits}}}) \quad \text{for} \quad i = 0, 1, \ldots, n_{\text{qubits}}-1.
\end{equation}
After entanglement, each qubit $q_i$ undergoes an additional rotation based on the trainable weight $w_{l,i}$:
\begin{equation}
    RY(w_{l,i}) = \exp\left(-i \frac{w_{l,i}}{2} \sigma_y\right).
\end{equation}

\textbf{3. Measurement:} The final stage involves measuring the states of the qubits. The expectation value of the Pauli-Z operator $\sigma_z$ is measured for each qubit, producing the output:
\begin{equation}
    \langle \sigma_z^i \rangle = \langle 0 | U^\dagger \sigma_z^i U | 0 \rangle,
\end{equation}
where $U$ represents the unitary operation of the quantum circuit. 

\vspace{-0.1in}
\subsection{Quantum Multimodal Contrastive Learning}
\vspace{-0.1in}
The multimodal data from the EEG and image encoders are fed into a contrastive learning framework. We use a cross-modal quantum circuit to entangle the quantum states of the two modalities. The similarity between corresponding EEG-image pairs is maximized while non-matching pairs are pushed apart. The framework uses a contrastive loss function that enforces these relationships across the quantum-encoded embeddings, ensuring that the learned representations capture the inherent multimodal correlations. The training process is conducted using a hybrid quantum-classical approach. The parameters of both the classical pre-processing modules and the quantum circuits are optimized jointly. We employ the parameter-shift rule for the quantum layers, while the classical parts are trained using gradient descent. The training objective is to minimize the contrastive loss, ensuring that the model learns to map corresponding EEG and image data to nearby points in the shared quantum embedding space.

\section{Result and Conclusion}
\vspace{-0.1in}
\subsection{Datasets, Preprocessing and Experiment Details}
\vspace{-0.1in}
The ThingsEEG dataset \cite{gifford2022large} contains extensive EEG recordings collected using a rapid serial visual presentation (RSVP) paradigm, capturing responses from 10 participants to 16,740 natural images from the THINGS database \cite{hebart2019things}. The dataset includes 1,654 training classes, each with 10 images, and 200 test classes, each with 1 image for zero-shot classification. EEG recordings were performed using 64-channel EASYCAP equipment, with data preprocessed by segmenting trials from 0 to 1,000 ms post-stimulus onset and applying baseline correction using the pre-stimulus mean. EEG responses for each image were averaged across repetitions. The images were resized to 224×224 pixels and normalized before further processing. All the subjects are trained by 200 epochs, we employ the weighted Adam optimizer, setting the learning rate to 0.0002 with \(\beta_1\) set to 0.5 and \(\beta_2\) to 0.999. The contrastive learning parameter \(\tau\) is initialized to \(\log(1/0.07)\). We use 10 qubits and 4 QNN layers for encoding. The results are averaged over 5 training runs using different random seeds.
\vspace{-0.1in}
\subsection{Result and Conclusion}
\vspace{-0.1in}
The experiment results are shown in TABLE~\ref{tab:qmcl_top1} and TABLE~\ref{tab:qmcl_top5}. The experimental results demonstrate that the QNervFormer-GA model outperforms the other models in both top-1 and top-5 validation accuracy across the 10 subjects in the ThingsEEG dataset. Specifically, QNervFormer-GA achieved the highest average top-1 accuracy at 5.2\%, closely followed by QNervFormer at 5.1\%. QSTConv-GA and QSTConv lagged behind with average top-1 accuracies of 4.0\% and 3.5\%, respectively. In terms of top-5 accuracy, QNervFormer-GA again led with an average of 20.3\%, slightly ahead of QNervFormer at 20.2\%. QSTConv-GA recorded an average of 16.8\%, while QSTConv was the lowest at 13.7\%. These results highlight the superior generalization ability of QNervFormer-GA, making it the most effective model among those tested for EEG-based image recognition on the ThingsEEG dataset.


\bibliographystyle{IEEEtran}
\bibliography{references,qml}

\end{document}